\documentclass[a4paper,12pt]{article}
\usepackage[utf8]{inputenc}
\usepackage{authblk}
\usepackage{rotating}
\usepackage[T1]{fontenc}
\usepackage{graphicx}
\usepackage{amsmath}
\usepackage{amsfonts}
\usepackage{amssymb}
\usepackage{xcolor}
\usepackage{soul}
\usepackage[colorlinks,citecolor=blue,linkcolor=blue]{hyperref}

\usepackage{float}
\usepackage{multicol}
\usepackage{booktabs}
\usepackage[numbers]{natbib}

\usepackage{flexisym}
\usepackage[font={sf},labelfont=bf]{caption}
\title{\bfseries In-situ observation of elastic instability of stress-induced B19$^\prime$ martensite in thin NiTi wires}
\author{Petr Sedl\'{a}k\textsuperscript{\dag}, Miroslav Frost\textsuperscript{\dag}, Martin \v{S}ev\v{c}\'i{}k\textsuperscript{\dag}, Luk\'{a}\v{s} Kade\v{r}\'{a}vek\textsuperscript{\ddag}, Hanu\v{s} Seiner\textsuperscript{\dag,*}}

\affil{\small \textsuperscript{\dag}{\it Institute of Thermomechanics, Czech Academy of Sciences, Prague, Czech Republic}}

\affil{\small \textsuperscript{\ddag}{\it FZU -- Institute of Physics, Czech Academy of Sciences, Prague, Czech Republic}}

\affil{{*}{corresponding author: hseiner@it.cas.cz}}

\begin{document}
\let\WriteBookmarks\relax
\def\floatpagepagefraction{1}
\def\textpagefraction{.001}

\maketitle

\begin{abstract}
A laser-ultrasonic approach was used to measure elastic properties of a superelastic nickel-titanium wire with the aim to evaluate their evolution with stress and temperature in stress-induced martensite. It was observed that this evolution can be well described by a single smooth surface in the stress-temperature space, with the values of Young's modulus ranging from 30 to 50 GPa. The evolution of the modulus was then monitored in-situ with further heating under fixed strain, that is, during the shape setting. The results revealed that the martensite phase experienced further softening during this process, reaching Young's modulus of nearly 10 GPa at high temperatures and high stresses. In addition, the measurement enabled a direct detection of the initiation and termination temperatures of the shape setting from the elasticity data, which was used to show that it occurs in the same temperature interval for tension-induced and torsion-induced martensite. 
\end{abstract}
\baselineskip16pt
\renewcommand{\arraystretch}{1.2}

\section{Introduction}

Polycrystalline  NiTi alloy is the most broadly used shape memory material nowadays, with applications ranging from biomedical engineering to space research \cite{Duerig_1999,Elahinia_2012,Hartl_2007,Biasutti_2024,Costanza_2020}.  
Elastic properties of this material are not only key parameters for designing these application, but their evolution when approaching the transition temperature also documents the gradual loss of mechanical stability of the lattice \cite{Nakanishi_1980}, enabling, thus, gaining a deeper insight into the process of the martensitic transformation (MT) itself {{\cite{NewRen}}}. This is well understood for the high-temperature cubic austenite, where the softening of elastic moduli prior to the forward transition temperature indicates the increasing instability with respect to the transformation path {{\cite{NewSittner}}}.

Similarly, the elastic stiffness of B19$^\prime$ martensite decreases with increasing temperature and this decrease accelerates close to the the MT temperature. This behavior has been documented in temperature cycles through the MT on both single-crystal {\mbox{\cite{Bodnarova2025,NewBrill}}} and polycrystalline {\mbox{{\cite{Grabec2021, NewDeng, NewBenafan, NewAhadi}}}} samples. At the same time,  a comparable elastic softening towards the MT has been documented for stress-induced MT in superelastic NiTi, where the elastic moduli of the martensite decrease with decreasing stress during unloading, until the reverse transition is triggered \cite{Sittner2014,Alonso2019}. Here, the interpretation of this elastic softening is not unambiguous, since increasing stress induces changes in the martensite microstructure and the measurements can be affected by inelastic mechanisms, such as twinning or de-twinning of martensite {\mbox{\cite{NewStebner, NewLiu, NewChowdhury}}}. The measurements were mainly performed by dynamical mechanical analysis (DMA) techniques, and the experimental results allowed for both interpretations: that the elasticity of the martensite increases with increasing stress, or that increasing stress reduces the inelastic mechanisms in the martensite, leading to an increase in its effective elasticity \cite{Alonso2019}. {Alternatively, the elastic properties of martensite were estimated from in situ diffraction measurements \mbox{\cite{NewQiu, NewStebner2}}, which allowed the elastic and inelastic processes to be separated.}

In this paper, we complement these observations using a unique in-situ ultrasonic measurement based on laser ultrasound, which probes the elastic response of the material with elastic waves with low stress amplitudes (below 1 MPa) and high frequencies (up to $\sim$2 MHz), which limits the possibility of any influence from inelastic processes in the microstructure. The results map the elastic behavior of martensite in a superelastic NiTi wire in a wide stress-temperature space, and cover also elevated temperatures and stresses, where martensite as a pure phase is no longer stable and transforms back to austenite by plastic deformation \cite{Sittner2018}. This allows us to discuss the elastic properties of martensite at high temperatures beyond its stability limit, and to demonstrate the used experimental approach to be a suitable tool for in-situ monitoring to the shape-setting process at these temperatures. 

\section{Materials and methods}

The experiments were performed on commercial high-strength superelastic medical-grade NiTi wires (straight annealed NiTi FWM\#1, {Fort Wayne Metals, USA)} with a diameter of 100 $\mu{}$m. The material behavior was described in detail in \cite{Sittner2009} within the Roundrobin SMA modeling activity, where comprehensive experimental data sets covering stress-strain responses at different temperatures, temperature cycling at different prestress levels, and recovery stress tests were collected. The main material parameters of the wire are summarized in table \ref{MaterPar}.

\begin{table}[ht]
\caption{Properties of the NiTi wire used for the experiments. }
\centering
\begin{tabular}{p{14cm}l}
\hline \hline
Material property& Value\\
\hline
Martensite start temperature & -72 $^\circ$C \\
Austenite finish temperature & 0 $^\circ$C \\
R-phase start temperature & 25 $^\circ$C \\
R-phase finish temperature & 5 $^\circ$C \\
Transformation stress of austenite at room temperature (RT, 24 $^\circ$C) &
555 MPa \\
Transformation strain of austenite at RT & 5.2\% \\
Yield stress at RT & 1400 MPa \\
Temperature change of the transformation stress (A$\rightarrow$M) &5.57 MPa/$^\circ$C \\ \hline \hline

\end{tabular}
\label{MaterPar}
\end{table}

For determining the elastic properties of the wire under the given stress-strain state and temperature, we utilized measurements of velocity of elastic wave propagation in the wire. We assumed that the wire behaves as a thin, elastically isotropic cylinder, with its elasticity described by two elastic constants: Young's modulus $E$ and shear modulus $G$. As discussed below, for characterization of tension-induced martensite we observed longitudinal and flexural modes of waves, that are both sensitive mainly to Young's modulus in the axial direction, which means that the possible elastic anisotropy of the wire resulting from its strong crystallographic texture {(documented for the given wire e.g. in {\cite{NewBian}} or {\cite{NewMolnarova}})} does not bias the results. For the twisting-induced martensite discussed in the final part of the paper, we observed torsional waves that are sensitive mainly to the shear stiffness along planes perpendicular for the wire axis, that is, also a single shear modulus $G$ is sufficient for the description.

\begin{figure}[!t]
\centering
\includegraphics[width=\textwidth]{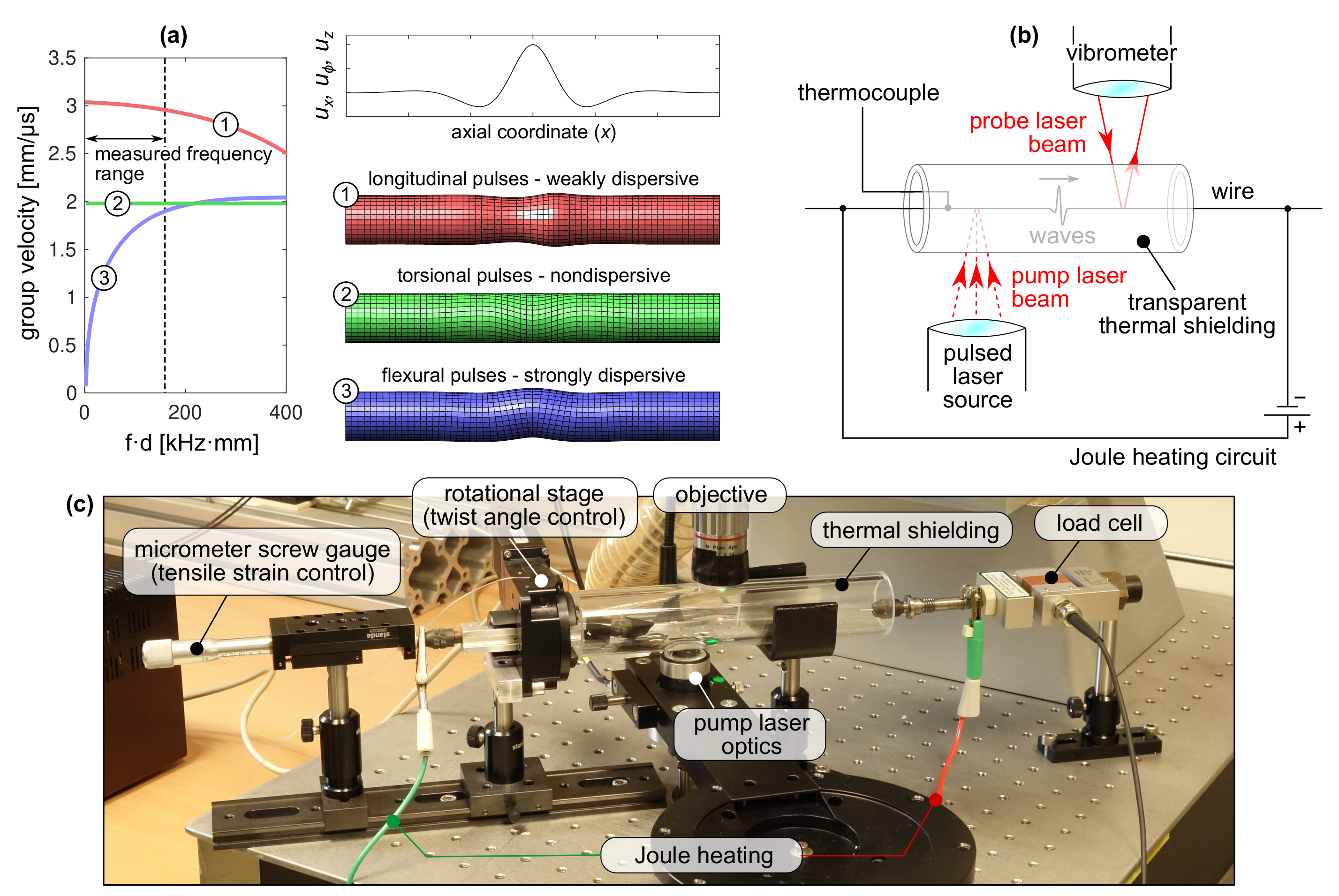}
\caption{Laser-ultrasonic characterization of NiTi wires: {\bfseries (a)} dispersion relations of the expected lowest-order modes. On the right, pulses of these modes are visualized as narrow wave packets of axial displacement ($u_x$), torsional displacement ($u_\phi$, where $\phi$ is the twist angle) and bending displacement ($u_z$, where $z\perp{}x$), respectively. {\bfseries (b)} a scheme of the measurement area with pump and probe lasers focused onto the wire. {\bfseries (c)} a photograph of the full set-up with external strain control and force measurement.}  \label{experiment}
\end{figure}

The wave propagation in the wire was studied using short laser pulses that were expected to initiate short acoustic pulses travelling along the wire. For a thin cylinder, three lowest-order modes of propagation can be expected to appear; longitudinal, torsional and flexural. In Figure \ref{experiment}(a), the frequency dependence of speed of propagation of short pulses (i.e., the group velocity) of such modes are shown, calculated for initially guessed elastic constants of NiTi, $E=$ 60 GPa, $G=$ 25 GPa. It is seen that the modes can be expected to experience very different dispersion behavior during the propagation, especially in the low-frequency range (calculation done using the Pochhammer equations \cite{Rose_textbook}). For the longitudinal and torsional modes, the velocity of propagation of the pulses exhibit very weak and zero frequency dependence, respectively. When initiated by pulse-like loading, these modes are expected to propagate as localized pulses, with small or zero broadening of the pulse. On contrary, the flexural mode is strongly dispersive, and thus, the initially short pulse of this mode is expected do develop into a broad wave packet during the propagation.

To detect these modes experimentally, a dedicated laser-source/laser-detection ultrasonic set-up was developed, utilizing the principle outlined in Figure \ref{experiment}(b). The wire was placed in a thick-walled PMMA tube that served for its thermal insulation from the ambient environment, and then its temperature was controlled using Joule heating and a thermocouple. Then, an infrared pulsed laser (nominal pulse width 8 ns, used pulse energy 1 mJ, wavelength 1064 nm,  Quantel ULTRA, USA) was focused through the PMMA tube onto the wire to generate ultrasonic pulses. The pulses were then detected in distance of 50 mm from the source by a laser-Doppler vibrometer (Polytec, Germany). The built-in microscope of the vibrometer was used to set the exact distance between the source and the detection point. By Fourier transform of the selected time-domain signals, it was concluded that the dynamic response of the wire was collected in a frequency range up to approximately 1.6 MHz (corresponding to $f\cdot{}d$= 160 kHz$\cdot$mm in Figure \ref{experiment}(a) for a 100 $\mu$m thin wire). For each temperature and each loading, the time-domain signals were obtained by averaging responses of 150 measurements with repetition rate of 5 Hz.

Figure \ref{experiment}(c) shows the experimental set-up with equipment for strain control, placed outside of the measurement area. The axial strain (elongation) of the wire was controlled by a screw gauge. In addition, a rotational stage enabled torsional loading by controlling the twist angle. The tensile-stress response to both types of loading was measured using a load cell (maximum force 20 N). The main aim of the experiment was to map the evolution of Young's modulus of the wire with temperature and tensile loading, both when the wire is fully in stress-induced martensite and when it undergoes temperature-induced irreversible transition (shape setting) to austenite under tensile stress. For all these measurements, the twist angle was set to zero. The rotational stage was used only for the experiment described in the final part of the paper, when the laser-ultrasound characterization was used to monitor the shape-setting process in a twisted wire. 

\begin{figure}
 \centering
 \includegraphics[width=\textwidth]{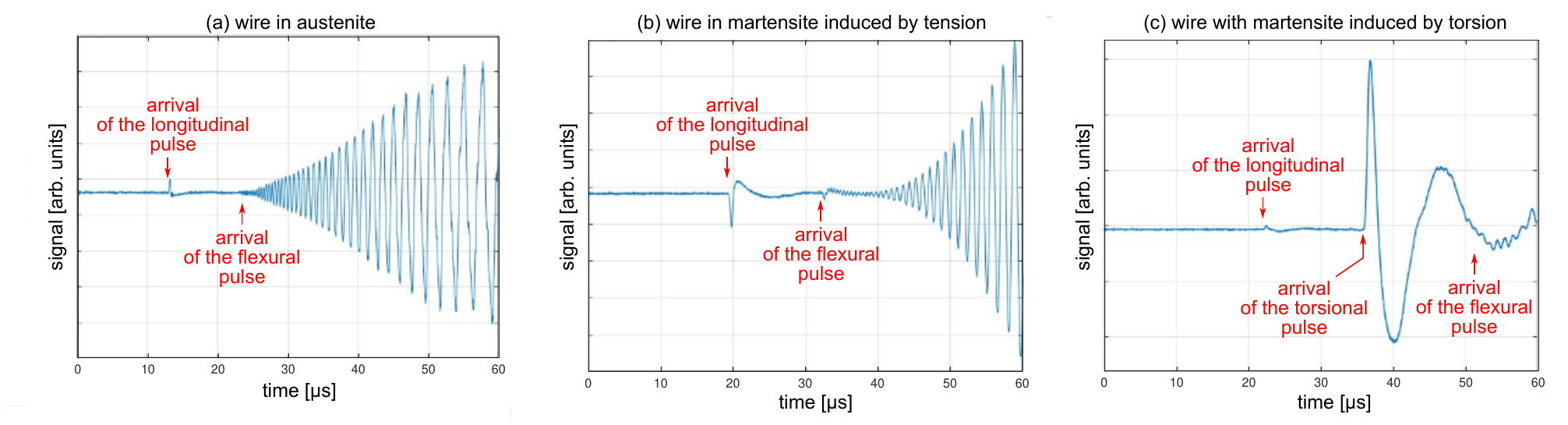}
 \caption{Time-domain signals recorded by the laser-Doppler vibrometer for different conditions of the wire. In addition to the longitudinal and flexural mode arrivals seen in (a) and (b), a dominant pulse corresponding to the torsional wave appears in (c).}
 \label{signaly}
\end{figure}

The elastic constants of the material were then calculated  from the arrival times of the sharp pulses, that is, the first longitudinal mode was used for $E=\rho{}v_L^2$, and the first torsional mode for $G=\rho{}v_T^2$, where $v_L$ and $v_T$ are the group velocities of the longitudinal and the torsional mode, respectively, and $\rho=6.46$ g$\cdot$cm$^{-3}$ is the mass density of NiTi.  The arrival times in both cases were understood as points at the footpoint of the pulse where the signal started to be distinguishable from the noise. For the longitudinal mode, which was weakly dispersive, this corresponded to the upper limit for the group velocity, that is, the limit reached for $f\cdot{}d\rightarrow{}0$ in Figure \ref{experiment}(a). Typical time-domain signals obtained in the measurement are shown in Figure \ref{signaly}, always exhibiting sharp pulses corresponding to longitudinal or torsional pulses, folowed by the arrival of the slower first flexural mode. The difference between the signals obtained for tension-induced and torsion-induced martensite will be discussed below.

\section{Experimental procedure and results}

\subsection{Young's modulus of stress-induced martensite}

\begin{figure}[!t]
 \centering
 \includegraphics[width=0.8\textwidth]{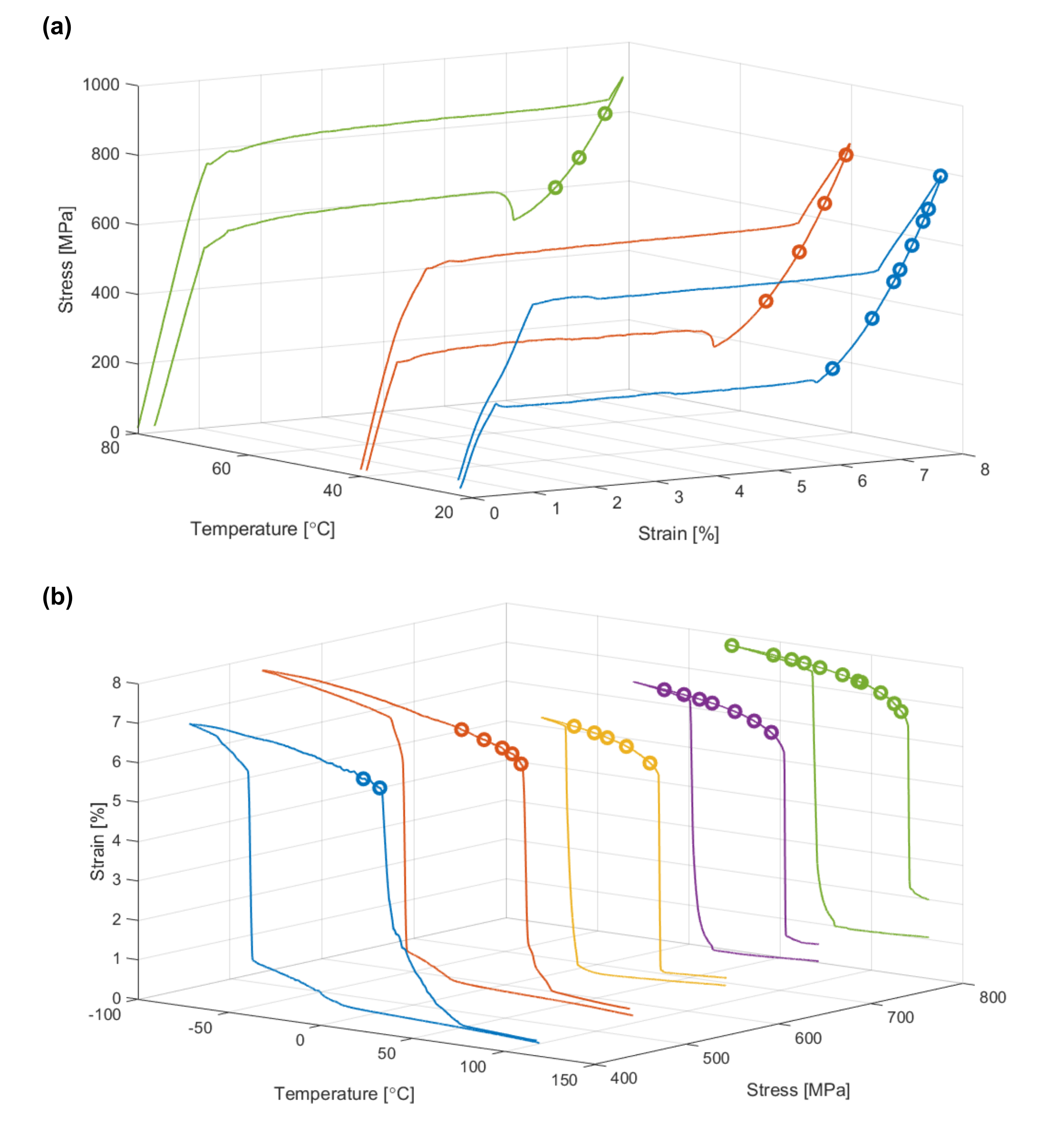}
 \caption{Stress-temperature conditions at which the Young's modulus of martensite was measured (a) in stress-strain cycles under constant temperature, (b) in temperature-strain cycles under constant stress.} \label{SS}
\end{figure}

The temperature dependence of elastic modulus of martensite in the superelastic range at conditions close to the reverse transformation was studied during tensile tests at constant temperature, and during thermal cycling tests at constant stress. In the first set of experiments, NiTi wire was strained up to 8\% of deformation at three constant temperatures (room temperature, 40 $^\circ$C and 80 $^\circ$C) to fully martensitic state, and then the Young’s modulus was recorded at several stress instances during unloading until the (localized) transformation back to austenite was initiated, see Figure \ref{SS}(a). In the second set of experiments, the wire was loaded until fully martensitic state was reached and then a constant stress level was maintained, and the elastic modulus was measured with heating at several temperature points preceding the reverse transformation. Such measurements were performed at stress levels from 400 MPa to 800 MPa with a 100 MPa increment, see Figure \ref{SS}(b). 

\begin{figure}
 \centering
 \includegraphics[width=0.8\textwidth]{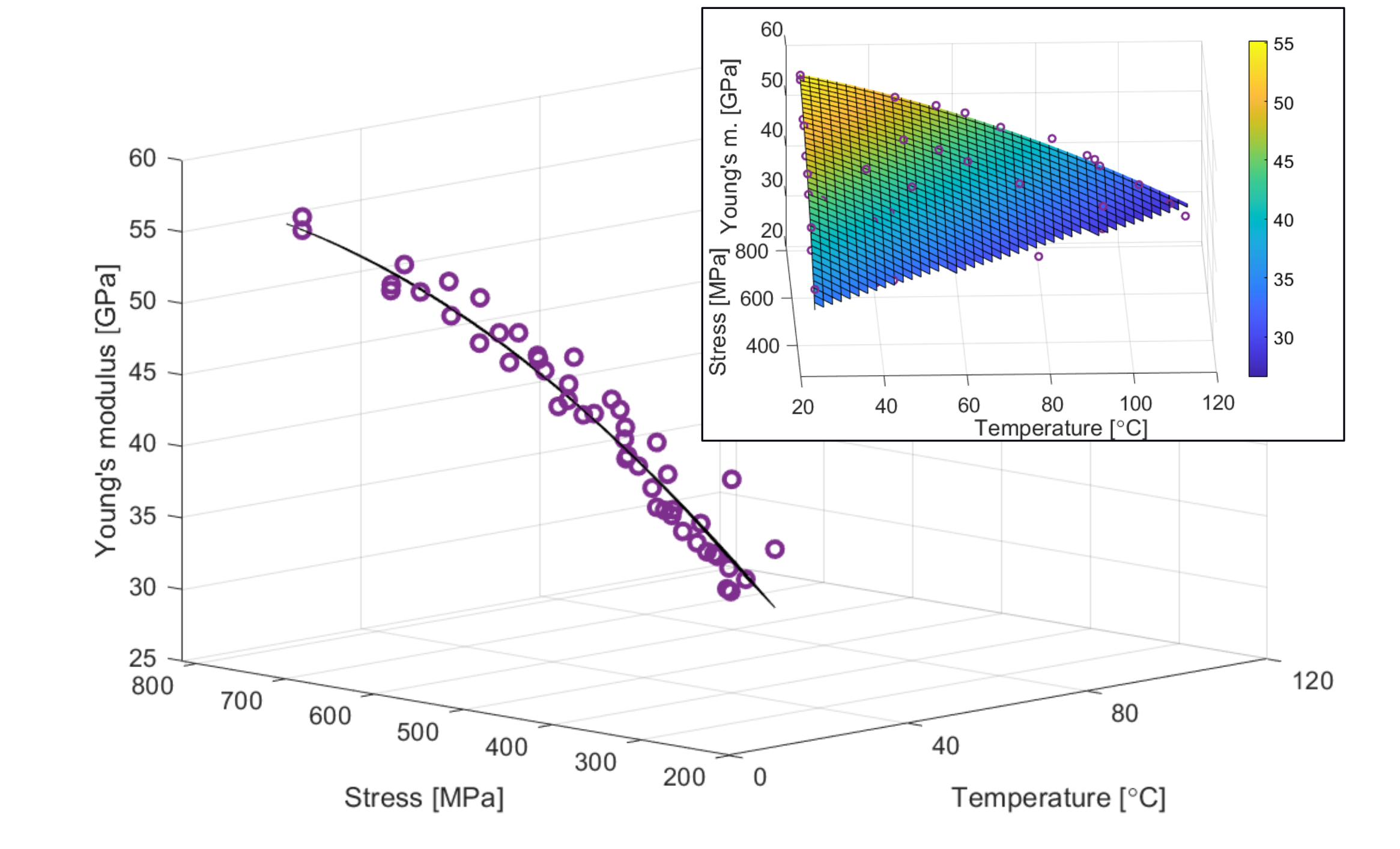}
 \caption{A Young's modulus evolution in the temperature-stress space approximated by a single smooth surface. The circles are the experimental data points, the smooth surface is a bivariate quadratic function fitting the experimental data in the least square sense. The color on the surface represents the value of $E$.} \label{surface}
\end{figure}

In this way, we obtained a set of points in the stress-temperature-Young’s modulus space. Inspired by \cite{Alonso2019}, we attempted to fit Young’s modulus values with a single function  depending solely on temperature and stress. The result of fitting of the experimental data by a (bivariate) quadratic function is a two-dimensional surface presented in the inset of Figure \ref{surface}. In the main plot of Figure \ref{surface}, the same 3D plot is rotated and tilted so that it is seen that the experimental data points match the surface very well. Apparently, the Young's modulus drops from values above 50 GPa (at the highest stress at room temperature) to less than 30 GPa when the transformation to austenite is approached, no matter which pathway is followed in the stress-temperature space (cf. \cite{Frost2010}). At the same time, the contours indicate that the Young's modulus changes only slightly along the stress-temperature points that are in the same distance from the martensitic transformation. According to the Clausius-Clapeyron relation, these points lie at lines parallel to the line corresponding to the start of the reverse martensitic transformation. 

\subsection{Elasticity measurement during shape setting under constant strain}

\begin{figure}[!t]
 \centering
 \includegraphics[width=0.6\textwidth]{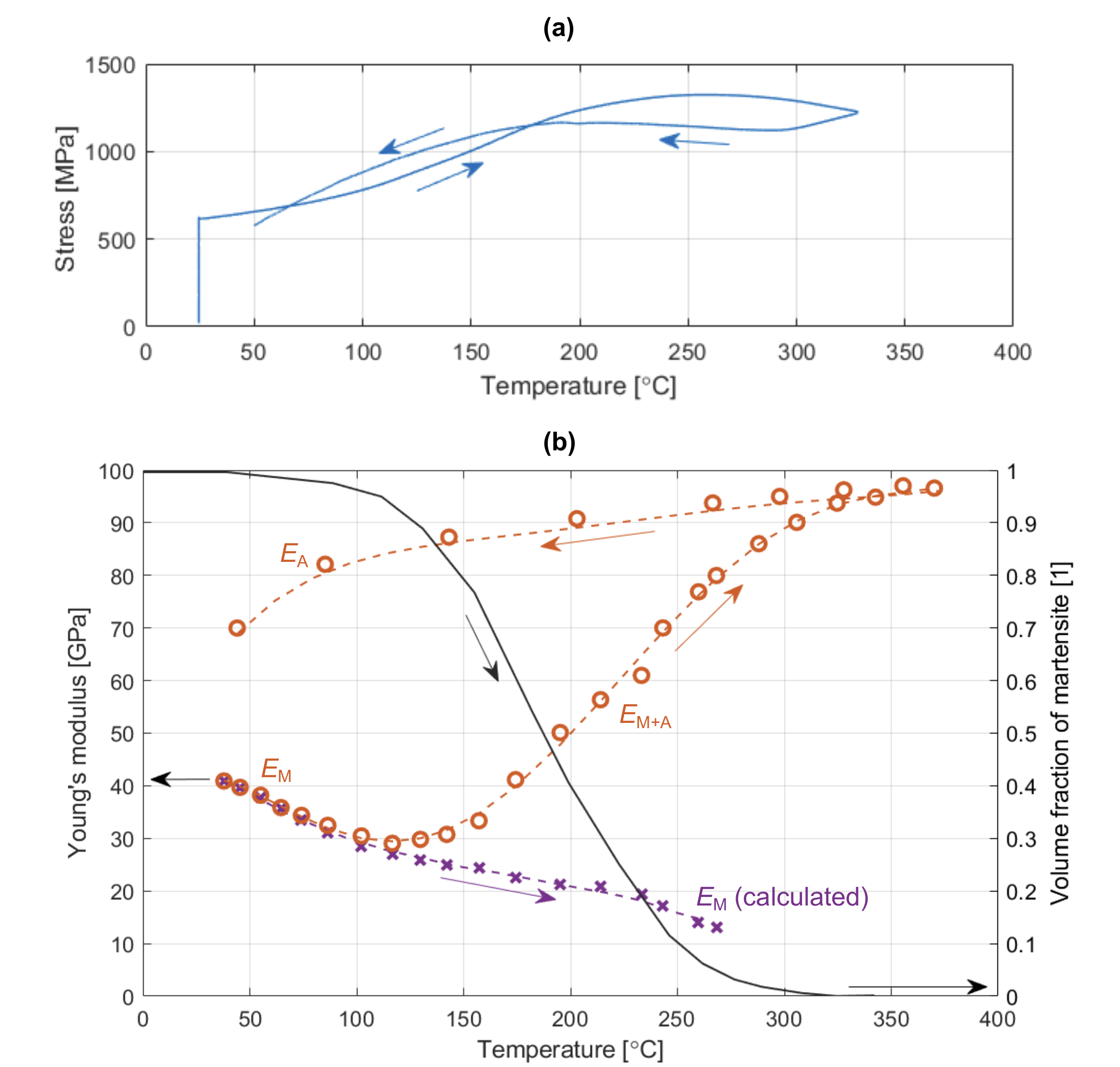}
 \caption{(a) the stress evolution in the wire with heating under fixed tensile strain; (b) the corresponding evolution of the Young's modulus. In (b) the red circles are experimental data obtained from the arrival times of longitudinal pulses, representing either elasticity of pure martensite ($E_{\rm M}$), or of pure austenite ($E_{\rm A}$), or of their mixture. The purple crosses are values    of $E_{\rm M}$ calculated from the behavior of the mixture. The solid black line shows the evolution of the volume fraction of martensite determined from X-ray diffraction experiments reported in \cite{Sittner2018}.} \label{LTSS}
\end{figure}

To directly monitor the temperature dependence of Young's modulus of martensite at the onset of the reverse transformation, we performed elasticity measurements during the so-called \emph{recovery stress test}. The wire was first loaded to the end of the plateau (7.2\% strain), which means it was fully in the stress-induced martensitic phase, and then heated under the constant strain. The wire was heated from room temperature (RT, 24 $^\circ$C) up to 370$^\circ$C and then cooled back to RT. The microstructural processes during this test were described in detail in \cite{Sittner2018, Heller2019}, we briefly summarize them here (see Figure \ref{LTSS}(a)): upon heating from RT, the stress in the wire increases in accordance with the Clausius-Clapeyron relationship. At approximately 100 $^\circ$C, the increase in stress with temperature still follows a linear trend, yet homogeneously distributed islands of plastically deformed austenite begin to appear in the martensitic matrix. {As seen in the X-ray diffraction data from {\cite{Sittner2018}} shown in Figure \mbox{\ref{LTSS}(b)}, }the volume fraction of austenite further increases with increasing temperature, and so does the stress, although the slope of the stress-temperature curve starts to decrease at temperatures at which about 50\% of the martensite has already transformed back to austenite. The reverse martensitic transformation is fully coupled with the generation of plastic deformation throughout the process, which must replace the transformation deformation of the oriented martensite and represents an unusual TRIP-like mechanism, called Low Temperature Shape Setting (LTSS) in {\cite{Sittner2018}}. At the temperature of approximately 300 $^\circ$C, the wire is completely transformed into austenite and remains in austenite with subsequent cooling, until reaching the temperature of 75 $^\circ$C, where it begins to transform into martensite because of the induced stress.

The results of the elasticity measurements during this process are plotted in Figure \ref{LTSS}(b)  by red circles. The initial heating of the martensite (up to 100 $^\circ$C) leads to a monotonic decrease in the Young's modulus. This result agrees well with the Young's modulus surface in Figure \ref{surface}, and the corresponding measured values lie on this surface. Further heating causes an increase in $E$, which starts deviating from the surface in Figure \ref{surface}. This is attributed to the nucleation of austenite and an increase in its volume fraction \emph{via} the LTSS mechanism. Above approximately 320 $^\circ$C, the modulus-temperature slope becomes constant, which indicates that the shape-setting process has been completed and the material is full in austenite. This is further confirmed during the cooling run, because above 320 $^\circ$C the heating and cooling curves perfectly coincide. With further cooling, the elastic modulus of austenite decreases with decreasing temperature, which is a well-documented behavior for metastable (entropically stabilized) high-temperature phases \cite{Nakanishi_1980}.

To estimate the evolution of $E$  of martensite above 100 $^\circ$C, where it occurs only in mixture with austenite, it was necessary to decouple the contributions of the individual phases to the overall elastic behavior. We used the results from synchrotron diffraction reported by \v{S}ittner et al. \cite{Sittner2018}, who determined the evolution of the martensite volume fraction during the LTSS process on the exactly same type of wires. We assumed that the elastic response of the mixture of austenite (Young's modulus $E_{\rm A}$) and martensite (Young's modulus $E_{\rm M}$) can be approximated by Hill's average \cite{Hill}
\begin{equation}
E_{\rm Hill}=\frac{1}{2}\left[\chi_{\rm A}{}E_{\rm A} + (1-\chi_{\rm A}) E_{\rm M} + \left(\chi_{\rm A}E^{-1}_{\rm A} + (1-\chi_{\rm A})E^{-1}_{\rm M}\right)^{-1}\right], 
\end{equation}
where $\chi_{\rm A}$ is the volume fraction of austenite, and for which we took $E_{\rm A}$ of austenite at given temperatures from the cooling curve. The resulting $E_{\rm M}$ is shown in Figure \ref{LTSS}(b) as purple crosses. It is seen that the elastic modulus of martensite inside of the mixture further monotonically decreases with increasing temperature. Let us point out that this decrease in $E_{\rm M}$ occurs simultaneously with the increase in tensile stress, which grows with heating according to the Clausius-Clapeyron rule (cf. Figure \ref{LTSS}(a)). As a result, a very low elastic modulus of martensite (below 20 GPa) is observed at stresses well above 1 GPa, where the martensite microstructure is already completely detwinned, and thus, any inelastic softening effect of martensite due to presence of twins can be clearly ruled out. Above $\sim$280 $^\circ$C, the volume fraction of martensite $(1-\chi_{\rm A})$ became too small to enable any reliable estimation of $E_{\rm M}$ from the Hill's homogenization. However, if we expect that  $E_{\rm M}$ follows the same trend as at lower temperatures and extrapolate this trend to 320 $^\circ$C, it turns out that the last islands of martensite probably disappear with Young's modulus well below 10 GPa; this agrees well with understanding the LTSS termination temperature as a temperature above which the martensite phase becomes thermodynamically unstable regardless of the applied stress.

\subsection{Elasticity measurement on a twisted wire}

The above results of laser-ultrasonic measurements during the LTSS process on a wire prestressed in tension (Figure \ref{LTSS}(b)) showed that the evolution of elasticity could be used to study the LTSS process itself, thanks to the large contrast between $E_{\rm A}$ and $E_{\rm M}$ at elevated temperatures. In particular, the start of LTSS can be easily identified from the deviation from the monotonic decrease in the effective Young's modulus of the wire with increasing temperature, and the end of LTSS corresponds to the temperature at which this modulus reaches that of austenite. This approach can be, thus also applied to study how does the LTSS process, which represents a reverse martensitic transformation coupled with plastic deformation, occur in martensite induced by a loading regime other than simple tension. In NiTi, it is well documented that the transformation strain of martensite changes with the loading regime, which is manifested, for example, by tension-compression asymmetry \cite{Liu_1998,Gall_1999,Bucsek2016}. Strongly $<$111$>$-textured NiTi wires show a large tension-compression asymmetry (the transformation strain in compression is almost half the transformation strain in tension) and the transformation strain in torsion is also much lower (around 60\% of the tension) \cite{Frost2015}. 

Different transformation strains result also in different thermomechanical-coupling coefficients: the Clausius-Clapeyron relationship implies that the increase in stress with temperature during the recovery stress tests is inversely proportional to the transformation strain, which means that the heating induces, for example, a much steeper increase in stress under fixed compressive strain than under fixed tensile strain. Consequently, a question arises, how do the different stress levels and different loading regimes (and different microstructures in stress-induced martensite in these regimes) affect the LTSS process. It is known that the plastic deformation mechanisms in NiTi are strongly affected by the microstructure \cite{Sittner2018}, and thus, the shape setting might be affected as well, as it runs \emph{via} irreversible, plastic mechanisms. 

In general, it is difficult to study the phase composition evolution during the shape setting on a single material under different loading regimes directly; however, the indirect tools such as the elasticity measurements can accompish this task. We performed the second LTSS experiment on the same wire, in which, however, the martensite was induced in torsion instead of tension. The wire was first twisted at RT to a twist angle of 2 Rad/mm, which corresponds to the end of the transformation plateau in torsion \cite{Sittner2009}. A small tensile prestress (50 MPa) was maintained in the axial direction to keep the wire straight.

X-ray diffraction tomography (DSCT) results reported in \cite{Sedlak2021} showed that under this condition the wire surface is fully in martensite, while the center of the wire remains is austenite. The volume fraction of martensite gradually decreases from the surface layer towards the center, giving rise to a core-shell structure in phase composition.
The same heating-cooling temperature cycle as in the LTSS tensile experiment was applied to the wire with a fixed twist angle. The ultrasonic signals detected during this experiment differed from those from tensile-loaded wires. The first sharp-edged signal detected corresponded again to the fastest longitudinal mode, but now a second sharp-edged signal corresponding to the torsional mode was also clearly detected, arriving before the dispersed wave packet of the flexural mode, and with an amplitude much higher than that of the longitudinal pulse (see Figure \ref{signaly}(c)).   

We rationalize this observation by the helical character of the martensitic microstructure that forms in wires in torsion \cite{Sedlak2021}, and that reacts to the rapid local heating from the pump laser by a torque pulse, unlike the axially symmetric microstructures that form in tension. Similarly, the torsional waves were not generated in pure austenite, where the grain microstructure is also axially symmetric.

The fact that arrivals of both the longitudinal and the torsional modes were detectable, allowed us to follow the evolution of two elastic constants during the heating: the Young's modulus $E$ directly related to the velocity of the longitudinal mode, and the shear modulus $G$ from the torsional mode.

\begin{figure}
 \centering
 \includegraphics[width=0.8\textwidth]{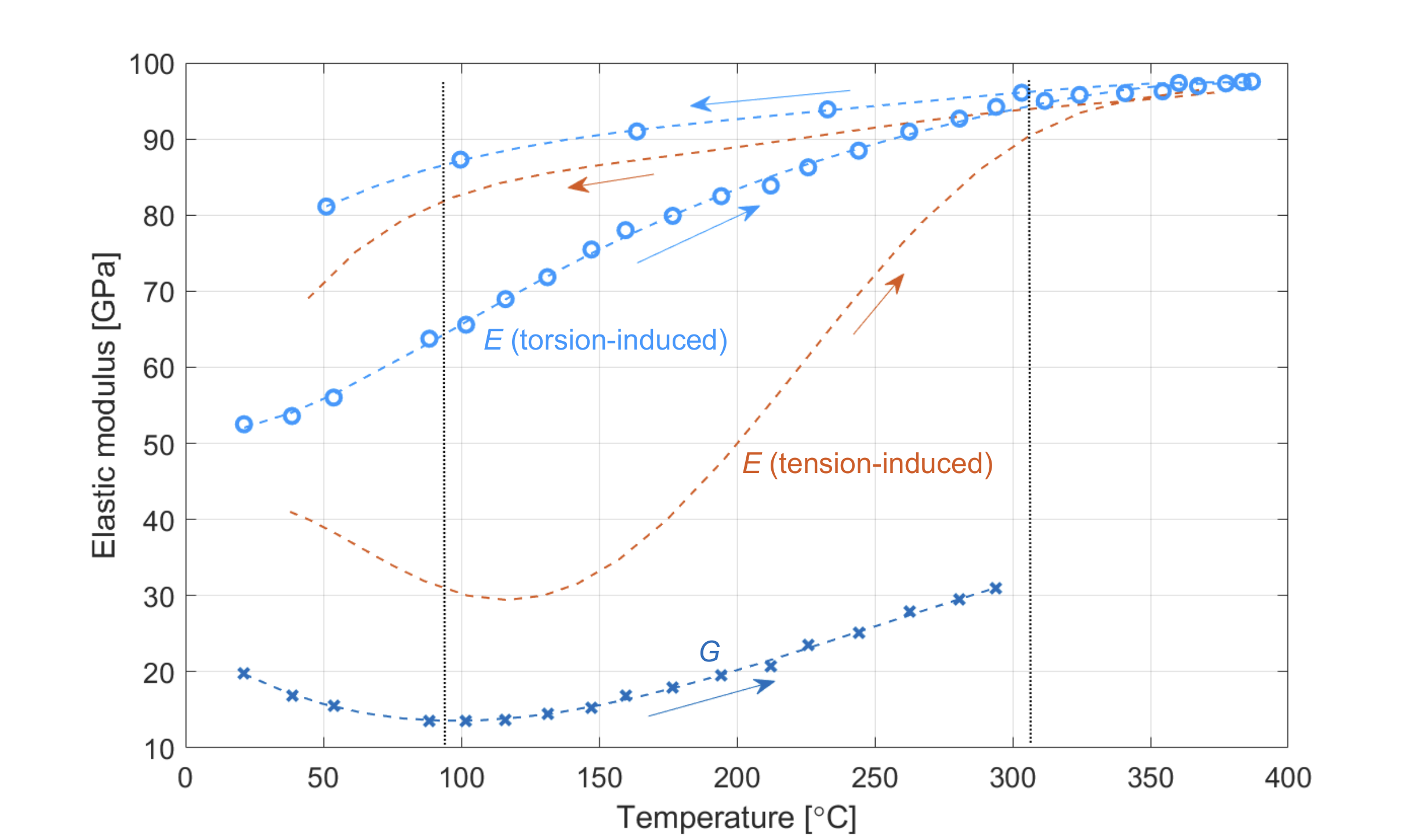}
 \caption{Evolutions of Young's modulus $E$ and shear modulus $G$ during the low-temperature shape setting cycle for martensite induced in torsion. The evolution in $G$ is shown only for the heating run, because the torsional pulse disappeared once the wire transformed fully into austenite. For comparison, the evolution of $E$ is shown also from the experiment with tension-induced martensite (the red dashes line).} \label{torsion}
\end{figure}

The results are plotted in Figure \ref{torsion}. The Young's modulus increased monotonically during the entire heating, unlike in tension. The absence of the initial softening part can be explained by two factors: first, the wire was not completely in martensite at the beginning, and the measured modulus $E$ averaged the responses from both the austenitic and martensitic parts, where the thermal hardening of austenite could dominate over the softening of martensite. Second, the large softening of the $E$ modulus before MT in martensite observed in wires loaded in tension was related to the transformation path and the loss of stability of the martensitic lattice with respect to the transformation strain. For martensite induced by twisting, the softening should be reflected in the shear modulus $G$, not in $E$. The evolution of $E$ was therefore not suitable for estimating the onset of the LTSS process. On the other hand, the point at approximately 310 $^\circ$C, where the $E$ curve became reversible and quantitatively reached the sane value as in the tension experiment, clearly marked the end of LTSS. 

The torsion-induced martensite can be expected to exhibit instability at high temperatures, and this instability should be, as mentioned above, related to the shear modulus. This is confirmed by the observed behavior of $G$, which is very similar to that of $E$ for tension-induced martensite, exhibiting first a decrease and than a continuous increase until the LTSS is completed. Note that the torsional mode has the highest amplitudes at the surface of the wire, and thus, its speed of propagation is quite insensitive to the elastic behavior of the austenitic core. For this reason, we can attribute the first evolution of $G$ fully to the changes of elastic behavior of surface layer of martensite. At approximately 90 $^\circ$C, the softening of $G$ changes into stiffening, signalizing the onset of the LTSS process and an increase in the austenite content on the wire surface. At approximately 300 $^\circ$C, the torsional mode disappears completely, indicating the disappearance of all martensite, and thus disappearance of any torque in the response of the wire to the pump laser pulse.

The reconstruction of the shear modulus of martensite from this experiment, as done above in tension for Young's modulus, is not possible. From the very beginning, the wire is in mixture of austenite and martensite, and we lack any information on the temperature evolution of the $G$ modulus of austenite (from the cooling part) due to the disappearance of the torsional mode. However, one important conclusion can be drawn from the results: the temperature interval of the LTSS process on the wire in torsion exactly corresponds to the temperature interval of the LTSS in tension.

The coincidence of both temperature intervals points to a fundamental finding on LTSS. It seems that this TRIP-like mechanism is dominantly driven by increasing temperature regardless of the development of external stress, or the morphology of the martensitic microstructure. In torsion, the transformation strain is significantly lower than in tension \cite{Frost2015}. Martensitic variants are thus accommodated into microstructures with lower macroscopic strain, and thus, a much higher (approximately 40\% higher)  equivalent stress is needed to maintain martensite as the stable phase during heating. However, this much higher stress does not cause easier plastic deformation processes, and LTSS in torsion is both initiated and completed at the same temperatures as in tension. Using the language of thermodynamics, this means that the dissipated energy during this process is the same regardless of the transformation strain of martensite. Similarly, the kinetics of the transformation of martensitic variants  to austenite  through plastic deformation is independent of the specific form of mechanical constraints (orientation and character of fixed strain) that prevent the reverse transition of martensite back to austenite via the classical transformation path. This experiment brings thus the first direct evidence of the distinctly non-Von Mises character of the LTSS process.

\section{Conclusion}

Experimental mapping of elastic behavior of stress-induced martensite reported here for a superelastic NiTi wire yielded two particularly interesting results. First, it was shown that Young's modulus of martensite decreases strongly towards the reverse martensitic transformation, regardless of whether the MT is approached by increasing temperature or by decreasing stress, or a combination of both. The Young's modulus of tension-induced martensite in the superelastic regime is described by a single surface in the stress-temperature space, and it value ranges from very low near the MT (around 30 GPa) to values above 50 GPa at high stresses or low temperatures. Along the M$\rightarrow$A transformation curve in the stress-temperature space, the elastic modulus of martensite decreases continuously with increasing temperature with a slope of around -5 GPa/100 $^\circ$C, and this decrease continues to elevated temperatures, when martensite as a pure phase is no longer stable and begins to transform to austenite through plastic deformation (the TRIP-like low-temperature shape-setting process, LTSS).

The increasing stability of austenite and the decreasing stability of martensite with further heating lead to a large difference between their elastic properties ($E_{\rm A}\approx{}10E_{\rm M}$ at 300 $^\circ$C). Here, this difference was exploited for in-situ monitoring of the LTSS process itself, which yielded the second important result: elasticity measurements during LTSS experiments on tension- and torsion-induced martensite showed that in both cases this heating-induced TRIP-like mechanism occurs in exactly the same temperature interval. Given the very different macroscopic transformation strains associated with the different loading regimes, this result is surprising and indicates the strongly non-Von Mises character of the generation of plastic strain during this process.

\section*{Acknowledgement}

This work has been financially supported by the Czech Science Foundation [project No. 25-16285S] and by the Operational Programme Johannes Amos Comenius of the Ministry of Education, Youth and Sport of the Czech Republic, within the frame of project Ferroic Multifunctionalities (FerrMion) [project No. CZ.02.01.01/00/22_008/0004591], co-funded by the European Union.

\section*{Data availability}

Data related to this article can be found at: https://zenodo.org/uploads/16793880

\end{document}